\documentclass{article}
\usepackage{graphicx} 
\usepackage{blindtext}
\usepackage{authblk}
\usepackage{setspace}
\usepackage{floatrow}
\usepackage{url}
\newfloatcommand{capbtabbox}{table}[][\FBwidth]
\usepackage[utf8]{inputenc}
\usepackage{blindtext}
\usepackage[a4paper, total={6in, 8in}]{geometry}
\usepackage{wrapfig}
\usepackage{lscape}
\usepackage{amsmath}
\usepackage{rotating}
\usepackage{epstopdf}
\usepackage{setspace}
\usepackage{subcaption}
\title{Modelling complexity in system safety: generalizing the $D^{2}T^{2}$ methodology}
\author[1]{Silvia Tolo}
\author[1]{John Andrews}
\date{}
\affil[1]{Resilience Engineering Research Group, University of Nottingham, U.K.}

\begin{document}

\maketitle
\begin{abstract}
    Although Fault Tree and Event Tree analysis are still today the standard approach to system safety analysis for many engineering sectors, these techniques lack the capabilities of fully capturing the realistic, dynamic behaviour of complex systems, which results in a dense network of dependencies at any level, i.e. between components, trains of components or subsystems. While these limitations are well recognised across both industry and academia, the shortage of alternative tools able to tackle such challenges while retaining the computational feasibility of the analysis keeps fuelling the long-lived success of Fault Tree and Event Tree modelling. 
    \\Analysts and regulators often rely on the use of conservative assumptions to mitigate the effect of oversimplifications associated with the use of such techniques. However, this results in the analysis output to be characterised by an unknown level of conservatism, with potential consequences on market competitiveness (i.e., over-conservatism) or safety (i.e., under-conservatism).
    \\This study proposes a generalization of the Dynamic and Dependent Tree Theory, which offers theoretical tools for the systematic integration of dependency modelling within the traditional Fault and Event Tree analysis framework. This is achieved by marrying the traditional combinatorial nature of failure analysis, formalised by the Fault and Event Tree language, with more flexible modelling solutions, which provide the flexibility required to capture complex system features. The main advantage of the proposed approach in comparison to existent solutions is the ability to take into account, under the same modelling framework, any type of dependency regardless of its nature and location, while retaining the familiarity and effectiveness of traditional safety modelling.
\end{abstract}

\section{Introduction}
\label{sec:intro}
The reliability and safety analysis of complex engineering systems generally relies on the gradual factorization of the overall system behaviour into that of its subsystems, trains of components and finally components. 
This is particularly evident in Fault Tree (FT) analysis \cite{ruijters2015fault} \cite{rauzy1993new}, and in their integration with Event Tree (ET) analysis \cite{vcepin2011event} within the framework of probabilistic safety assessment \cite{andrews2002reliability} \cite{stamatelatos2000probabilistic}.
Such a framework can be split into three main modelling levels:
\begin{itemize}
    \item Basic events (BEs), associated with the components mathematical failure models;
    \item Fault trees, detailing the failure mechanisms internal to subsystems;
    \item Event trees, representing the interaction of subsystems failure events and the resulting consequences on the overall system;
\end{itemize}
Each of these modelling layers feed into each other, with the component failure models becoming the input of subsystems FTs in the form of basic event probabilities, and the analysis of the subsystems FTs informing the branches of the ET, from which the overall system response to specific accident scenario can be finally estimated.
\\Although this approach has been at the core of reliability science in most engineering sectors pretty much from its introduction, dating back to the mid 1960s, it is characterised by a significant limitation spreading across all modelling levels: the representation of the dynamic behaviour of systems and of the resulting dependencies existing among its parts \cite{kabir2017overview}. 
\\The generic definition of dependency embraces any kind of relationship between two or more events which results in the reciprocal influence over the likelihood of their occurrence \cite{zeng2023dependent}. Two events are considered directly dependent if knowing the state of one (e.g., occurred or not occurred) changes the probability of the other event to occur. The dependency between two events can also be indirect, and hence be triggered by a further event or chain of events, on which both depend on directly (e.g., in the case of shared resources). A closed set of events connected to each other by direct or indirect dependencies, while independent from any other events within the same model, is referred to as a \textit{dependency group}.
\\The behaviour of any real-world system is in reality characterized by a dense network of dependencies, in many cases exacerbated by the heavy employment of automation and control technology common to modern systems \cite{meango2019failure} \cite{fricks1997modeling}. Dependencies can be associated with any degree of system complexity: i.e., between individual components (e.g., secondary failures, maintenance), trains of components within subsystems (e.g., redundancy) or between the subsystems themselves (e.g., causal dependencies, shared resources). Hence, in order to predict adequately the realistic failure behaviour of systems, it is necessary to implement approaches able to systematically model complex relationships existing at any level of the physical system.
\\The Dynamic and Dependent Tree Theory, also known as $D^{2}T^{2}$, aims at offering solutions for integrating any type of dependencies across the traditional FT/ET modelling framework \cite{ANDREWS2023108959}. 
This is achieved through the use of flexible modelling solutions, such as Markov Models (MMs) \cite{ gagniuc2017markov}, Petri Nets (PNs) \cite{reisig2012petri}, AltaRica \cite{arnold1999altarica} etc. The use of these is tailored on modelling only relationships existing between dependent events, allowing to preserve the strengths of traditional solutions, such as the familiarity and efficiency of the FT/ET language.
\\Differently from existent techniques, such as Dynamic Fault trees \cite{vcepin2002dynamic}\cite{dugan1992dynamic}\cite{aslansefat2020dynamic}  Boolean logic Driven Markov Processes \cite{bouissou2003new} and Pandora temporal FTs \cite{walker2007compositional}, the $D^{2}T^{2}$ doesn't require the use of predefined templates (e.g., dynamic gates), allowing to depict dependency relationships regardless of their nature, location, or modelling level involved.
\\The following sections offer an overview of the theoretical background of the methodology (Section \ref{sec:method}) and describe in detail the strategy proposed to tackle the modelling of dependency across all layers of traditional system safety analysis, from basic component dependencies (Section \ref{sec:basicDep}), to trains of components (Section \ref{sec:intermediateDep}) as well as subsystems dependencies (Section \ref{sec:etDep}).

\section{Theoretical Background}
\label{sec:method}
The $D^{2}T^{2}$ methodology offers a generic approach to model dependencies occurring across any modelling level of traditional FT/ET analysis. This includes dependencies within individual FTs, either between basic (referred to as \textit{basic dependencies}) or intermediate events such as FT gates (namely \textit{intermediate dependencies}), as well as relationships existing between multiple subsystems and hence FTs. The strength of the dependency relationships is captured numerically through the use of joint reliability metrics (i.e., joint probabilities, frequencies) and the approach strongly relies on their mathematical manipulation, as discussed in Section \ref{sec:jointP}. Joint metrics are then re-integrated in the overall modelling framework through the use of Binary Decision Diagrams \cite{rauzy1993new} \cite{sinnamon1996fault} and the adoption of novel algorithms, as briefly introduced in Section \ref{sec:BDD}. 

\subsection{Dependency Representation}
\label{sec:jointP}
The algorithms developed within the $D^{2}T^{2}$ framework rely on the use and manipulation of joint reliability metrics. These are adopted to quantify the dependency relationships underlying two or more events, and can be gathered directly from experimental data when available, or from the analysis of support models mimicking the dependency mechanisms, e.g., PNs or MMs. For ease of understanding, the focus of the current discussion is restricted to the case of joint probabilities, however the principle presented can be easily extended to frequencies.
\\Given two dependent Boolean events $\mathbf{A}=\{A,\overline{A}\}$ and $\mathbf{B}=\{B,\overline{B}\}$, their dependency can be represented as the set of probabilities associated with each possible combination of states of the two events. Such set consists of:
\begin{itemize}
    \item $q(A,B)$: probability associated with the simultaneous occurrence of both events $\mathbf{A}$ and $\mathbf{B}$;
    \item $q(A,\overline{B})$: probability of only $\mathbf{A}$ occurring;
    \item $q(\overline{A},B)$: probability of only $\mathbf{B}$ occurring;
    \item $q(\overline{A},\overline{B})$: probability of both $\mathbf{A}$ and $\mathbf{B}$ not to occur;
\end{itemize}
Further information can be gathered about the likelihood of the dependent events through the use of basic joint probability manipulation rules, such as \textit{marginalization} and \textit{conditioning}.
\\The first allows to estimate the probability of a sub-set of the events embraced in the original dependency group, simply summing together all joint probability values including the sub-set of interest. For instance, the overall probability of event $\mathbf{A}$ to have occurred (i.e., $\mathbf{A}=A$), can be calculated as:
\begin{equation}
    q(A)=q(A,B)+q(A,\overline{B})
\end{equation}
hence considering all joint events in the initial set which include the state $A$ regardless of the state of $\mathbf{B}$. 
\\The conditional probability of the remaining event $\mathbf{B}$ to not occur, given that $\mathbf{A}$ has occurred (i.e., $q(\overline{B}|A)$), can instead be obtained through conditioning as:
\begin{equation}
    q(\overline{B}|A)=\frac{q(A,\overline{B})}{q(A)}
\end{equation}
\\Dependency structures can be quite complex, involving multiple events and also indirect dependencies. These refer to the case in which two or more events are dependent on each other through a third source. For instance, let's consider a third event $\mathbf{C}$, also directly dependent on $\mathbf{B}$. Since the probability associated with the occurrence or non-occurrence of $\mathbf{B}$ changes according to the state of $\mathbf{C}$, which in turn affects the probability of $\mathbf{A}$, then all three events $\mathbf{A}$, $\mathbf{B}$ and $\mathbf{C}$ are related to each other, in spite of $\mathbf{A}$ and $\mathbf{C}$ not having any causality relationship. This establishes an indirect dependency between the two events, so that:
\begin{equation}
    q(\mathbf{B},\mathbf{C})\neq q(\mathbf{B})q(\mathbf{C})
\end{equation}
 Indirect dependencies such this can be removed through \textit{instantiation}: when the state of the dependency source is known, the two indirectly dependent events become independent. This solution relies on the principle of \textit{D-separation} \cite{geiger1990d}, which consists of 'blocking' the dependency through instantiating the source of such dependency.  
 \\With regards to the previous example, if the state of $\mathbf{B}$ is known, then $\mathbf{A}$ and $\mathbf{C}$ become independent on each other, such that:
\begin{equation}
    \begin{split}
q(\mathbf{A},\mathbf{C}|B)&=q(\mathbf{A}|B)q(\mathbf{C}|B)\\  q(\mathbf{A},\mathbf{C}|\overline{B})&=q(\mathbf{A}|\overline{B})q(\mathbf{C}|\overline{B})
    \end{split}
\end{equation}

\subsection{Binary Decision Diagrams}
\label{sec:BDD}
\begin{figure}
     \centering
     \includegraphics[width=0.8\linewidth]{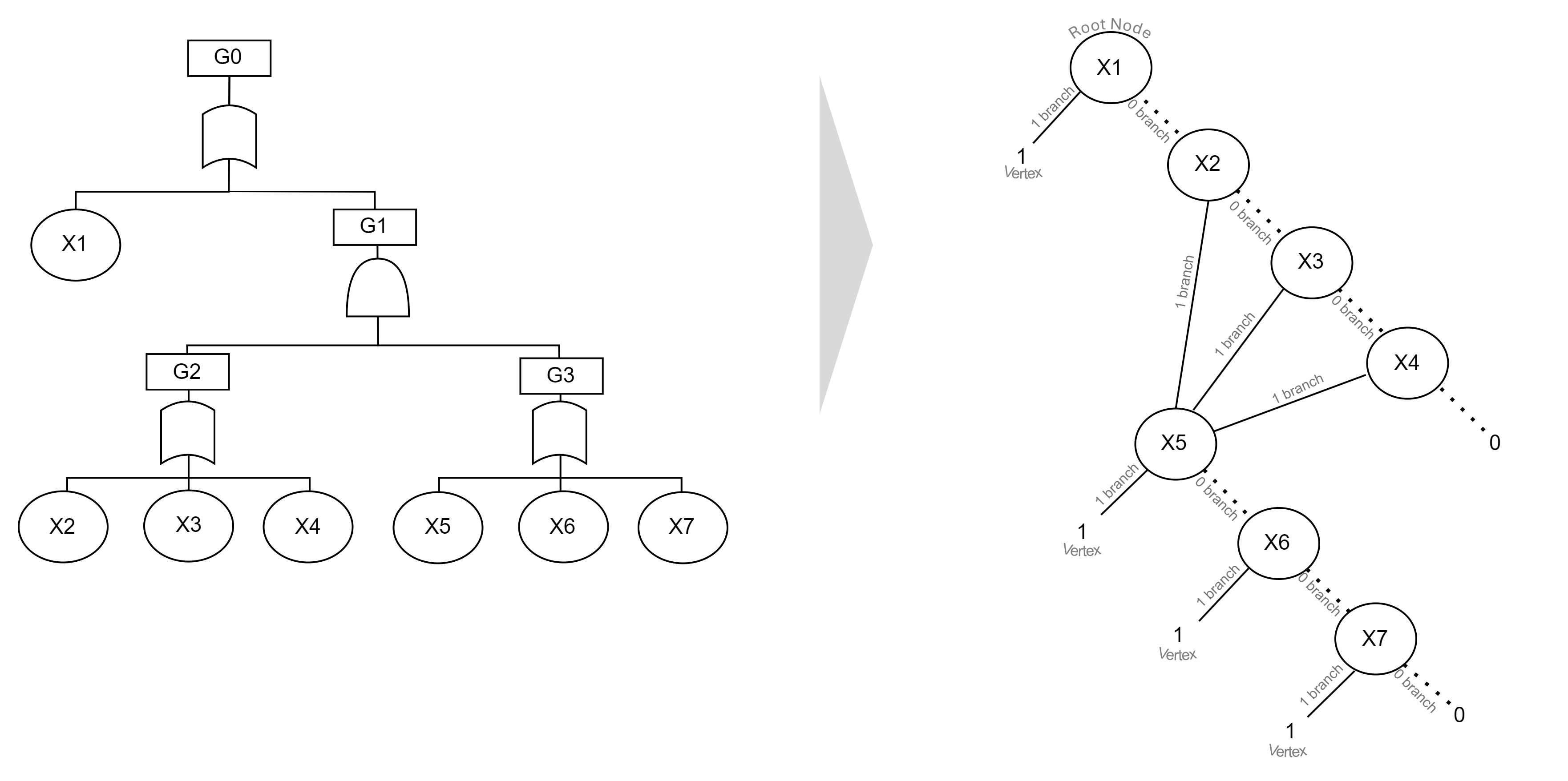}
     \caption{Example of FT/BDD conversion}
     \label{fig:bddft}
 \end{figure}

 BDDs are widely adopted in current academic and industrial practice for expressing the Boolean function econded by FTs in a more efficient format, enhancing the computational performance of FT analysis. As shown in Fig.\ref{fig:bddft}, they consists of nodes, edges and vertices: the first correspond to FT basic events, and are usually labelled accordingly. From each node originates two edges: one referring to the occurrence of the associated event, referred to as the 1 branch, the other to the non-occurrence of the same (i.e., 0 branch). Finally the terminal vertices of the graph indicate the state of the encoded FT top event: by convention, vertex 0 refers to the non-occurrence of the top event, 1 to its occurrence.
 \\Any combination of edges linking the BDD root node to a vertex is referred as a path, and describes a combination of basic events outcomes which result in the occurrence of top event, if ending with vertex 1, or its not-occurrence if ending in a vertex 0. 
In light of this, the computation of the corresponding FT top event reliability metrics can be attained through the identification of all failure paths linking the root node to a terminal vertex one ($path_i$), summing the probability associated to each path. This can be expressed as:
\begin{equation}
    Q(TOP)=\sum_i q(path_i)
\end{equation}
In the absence of dependencies, failure path probabilities can be calculated as the product of the probability associated with each individual event included in the path, such as:
\begin{equation}
    q(path_i)=\prod_{j|Xj\in path_i} q(X_j)
\end{equation}
However, if dependent events are included in the path, their joint probability must be taken into account in place of marginal probabilities. This can be expressed as:
\begin{equation}
\label{eq:pathP}
    q(path_k)= q(path_k)^{DG0} \prod_{z=1}^{n} q(path_k)^{DGz}
\end{equation}
where the label $DG$ indicate dependency groups, namely closed sets of events sharing a certain dependency relationship. According to the $D^2T^2$ convention, the dependency group labelled $DG0$ includes independent events, so that $q(path_k)^{DG0}$ in Eq.\ref{eq:pathP} indicates the contribution to the overall path probability of the independent events. The contribution of the remaining dependency groups ($q(path_k)^{DGz}$ in Eq.\ref{eq:pathP}) relies on the use of joint probabilities associated with the specific combination of states of the dependent events belonging to the path. If the path includes only a subset of the overall dependency group, the manipulation of the original joint probability through marginalization may be required, as discussed in Section \ref{sec:jointP}.
A detailed algorithm for the computation of BDDs in the presence of dependencies is provided in \cite{tolo2023fault}.
\\Beyond the top event reliability metrics, the BDD computation can also target the computation of conditional and joint reliability. 
This is achieved selecting the failure paths appropriately: the probability of the top event to occur together with a certain basic event $X_h$, can be calculated summing only the probability of failure paths compatible with $X_h$ itself. For instance, with reference to the example model in Fig.\ref{fig:bddft}, the probability of the top event $TOP$ to occur together with basic event $X2$ would be equal to:
\begin{equation}
    Q(TOP,X2)=\sum_{i|\overline{X2} \notin path_i} q(path_i)
\end{equation}
This imply the summation to be calculated over four paths, such as:
\begin{itemize}
    \item ${X1}$
    \item ${\overline{X1},X2,X5}$
    \item ${\overline{X1},X2,\overline{X5},X6}$
    \item ${\overline{X1},X2,\overline{X5},\overline{X6},X7}$
\end{itemize}
since all remaining paths would include the complimentary event of $X2$, $\overline{X2}$.
\\Conditional probabilities can be obtained following a similar procedure, but excluding $X2$ from the path probability calculation. With reference to the previous example, this can be expressed as:
\begin{equation}
    Q(TOP,X2)=\sum_{i|\overline{X2} \notin path_i} q(path_i|X2)
\end{equation}
where the conditional probability for X2 can be calculated from the same paths listed above upon removal of $X2$.
\\Once the paths relevant to the metrics of interest are identified, the procedure to follow for the calculation of their individual probability will depend on the presence or lack of dependent events within the same paths.

\section{Methodology}
The $D^2T^2$ offers a systematic approach to integrate dependencies within any level of the FT/ET modelling framework. As discussed in the previous section, this is achieved through the use of joint probabilities and their integration within BDDs computation. However, while these solutions stand as the foundation of the approach regardless of the dependency type, their implementation depends on the associated modelling layer, requiring different solutions. 
\\The following sections focus on presenting the $D^2T^2$ procedure dedicated to tackle dependencies between FT basic events (Section \ref{sec:basicDep}), between intermediate events (Sections \ref{sec:intermediateDep}) and finally between entire FTs, as investigated in Section \ref{sec:etDep}.
\begin{figure}
\begin{subfigure}{.5\textwidth}
  \centering
  \includegraphics[width=\linewidth]{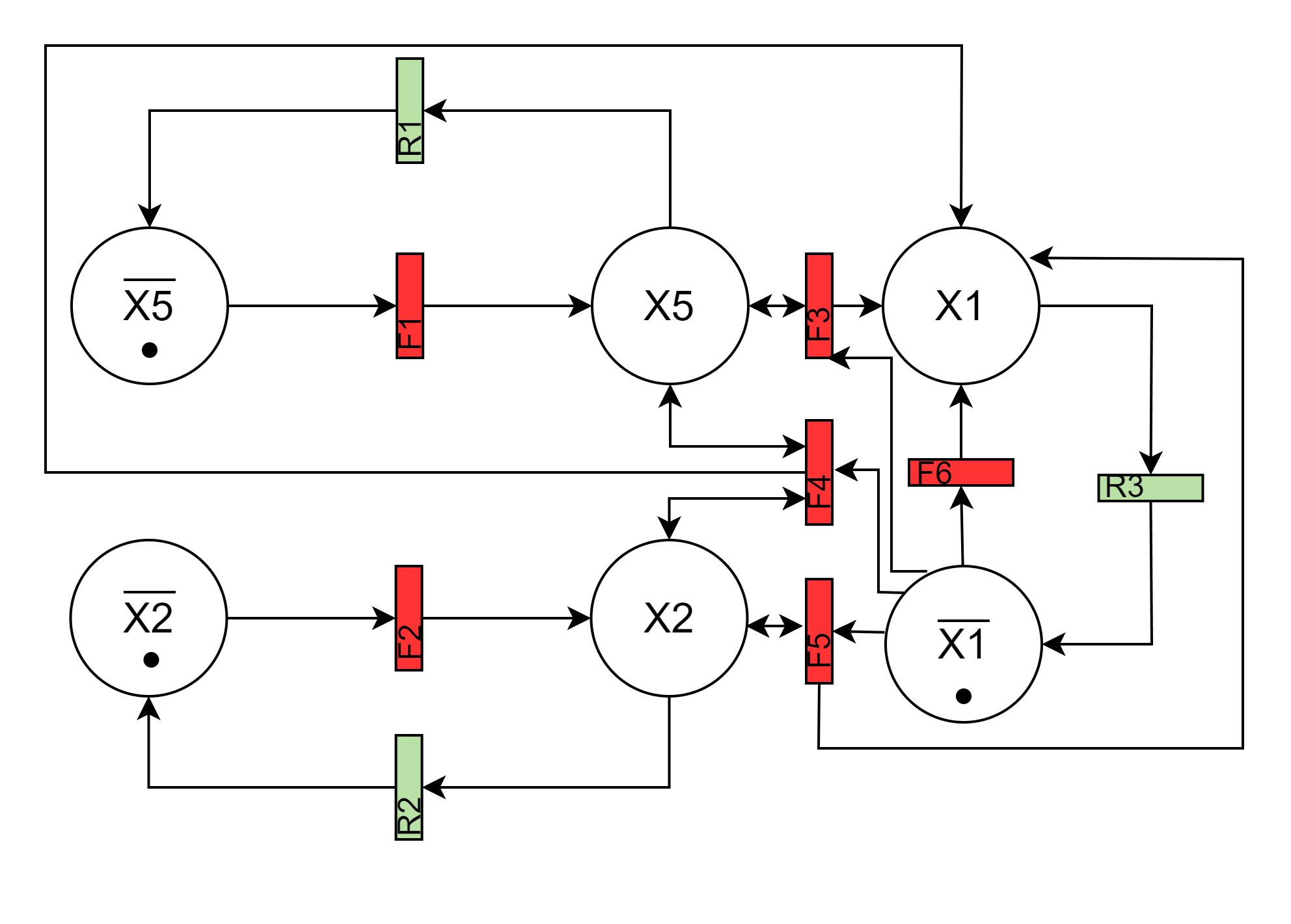}
  \caption{Example PN}
  \label{fig:BEpn}
\end{subfigure}%
\begin{subfigure}{.5\textwidth}
  \centering
  \includegraphics[width=\linewidth]{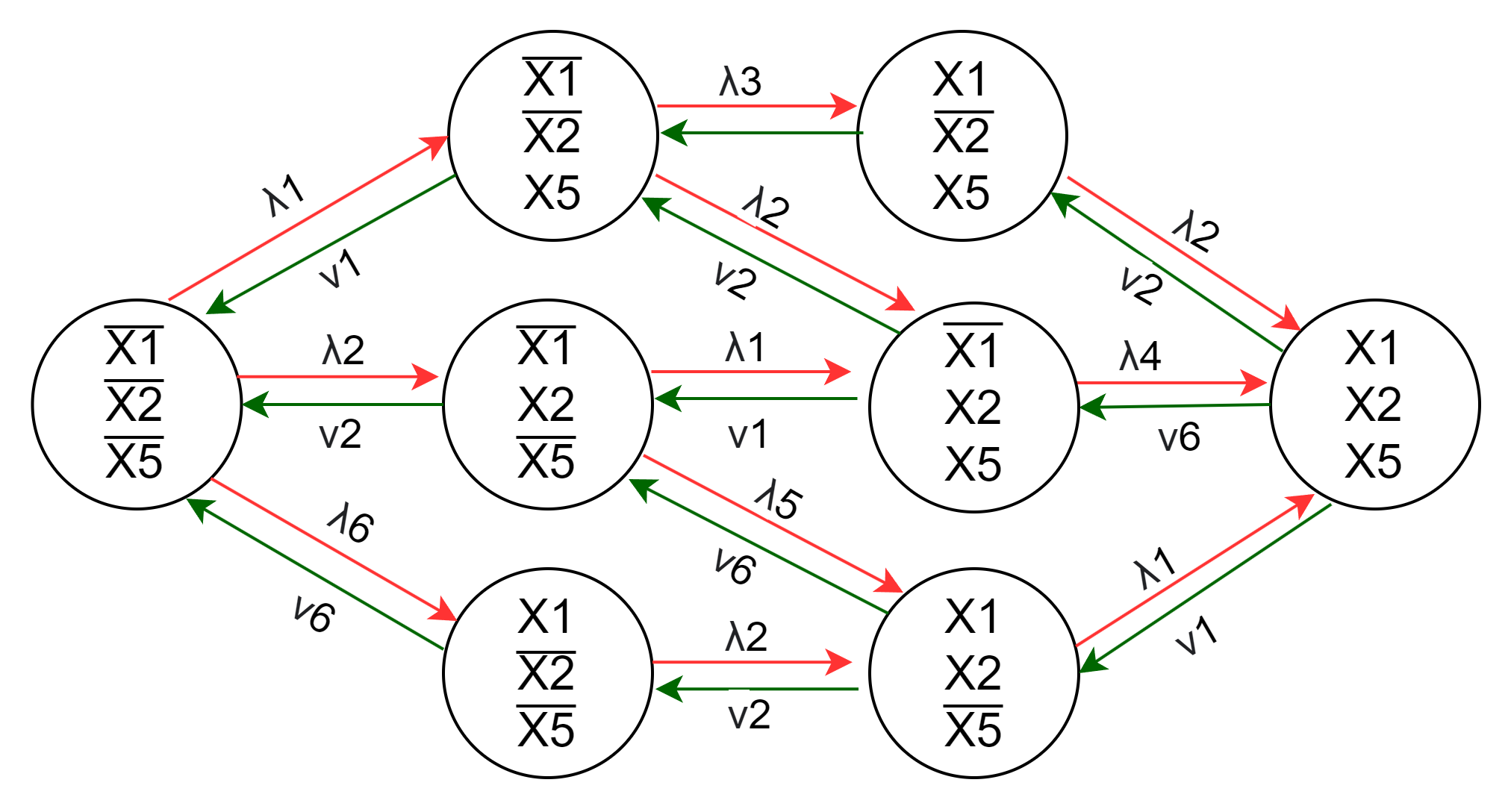}
  \caption{Example MM}
  \label{fig:BEmm}
\end{subfigure}
\caption{Example models for basic dependencies}
\label{fig:fig}
\end{figure}

\subsection{Basic Events Dependency}
\label{sec:basicDep}
\begin{figure}
\begin{floatrow}
\ffigbox{%
  \includegraphics[width=.7\linewidth]{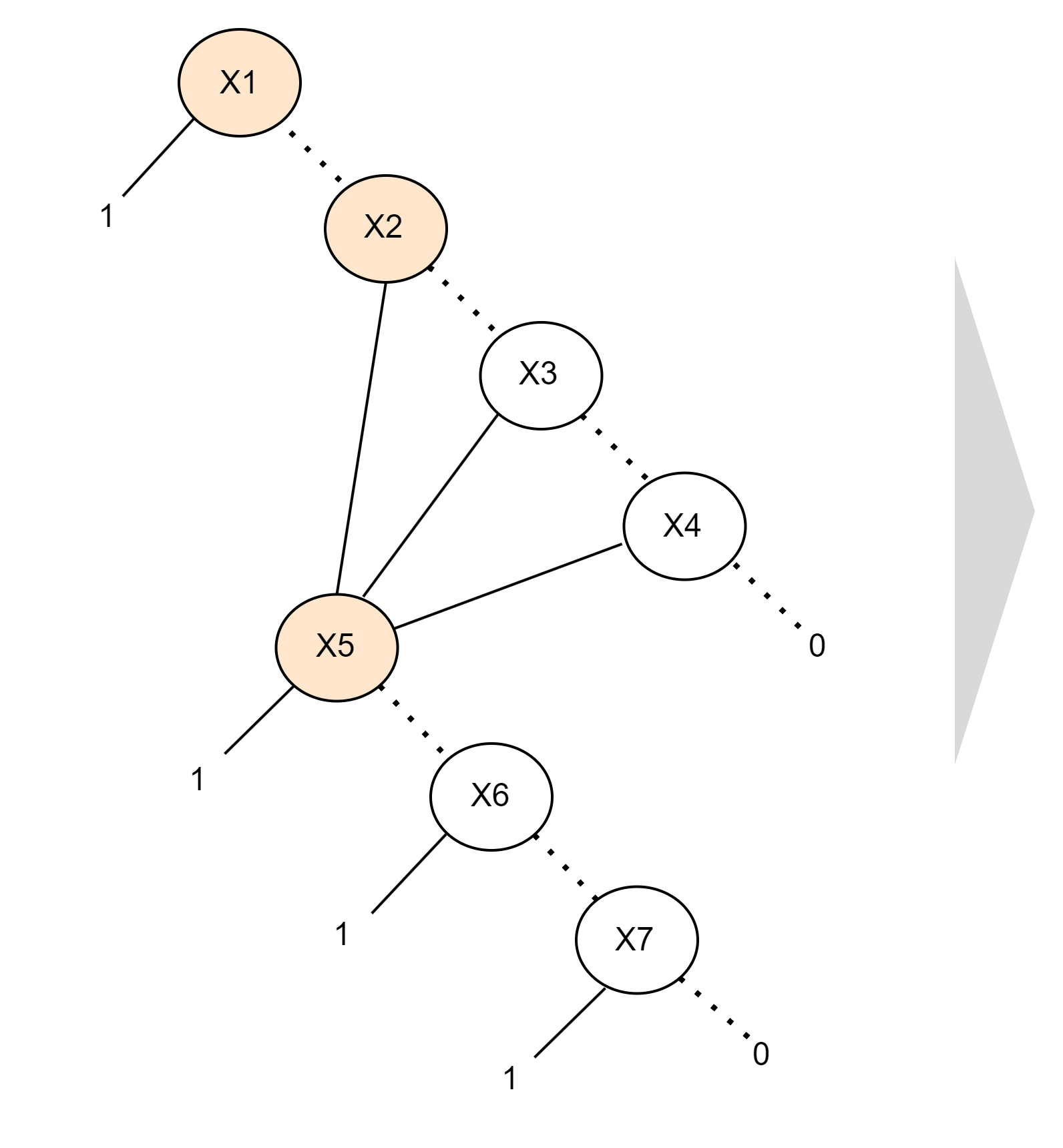}
}{%
  \caption{BDD with dependency group DG1}%
}
\capbtabbox{%
   
  \begin{tabular}{cll} \hline
  PAth ID & q(DG0) & q(DG1)\\ \hline
  1 &   & $q(X1)$   \\
  2 &  & $q(\overline{X1}, X2, X5)$  \\
  3 & $q(X3)$  & $q(\overline{X1}, \overline{X2}, X5)$  \\
  4 & $q(\overline{X3})q(X4)$  & $q(\overline{X1}, \overline{X2}, X5)$  \\
  5 & $q(X6)$ & $q(\overline{X1}, X2, \overline{X5})$  \\
  6 & $q(X3)q(X6)$  & $q(\overline{X1}, \overline{X2}, \overline{X5})$  \\
  7 & $q(\overline{X3})q(X4)q(X6)$  & $q(\overline{X1}, \overline{X2}, \overline{X5})$  \\
  8 & $q(\overline{X6})q(X7)$ & $q(\overline{X1}, X2, \overline{X5})$  \\
  9 & $q(X3)q(\overline{X6})q(X7)$  & $q(\overline{X1}, \overline{X2}, \overline{X5})$  \\
  10& $q(\overline{X3})q(X4)q(\overline{X6})q(X7)$  & $q(\overline{X1}, \overline{X2}, \overline{X5})$  \\
  \hline
  \end{tabular}
}{%
  \caption{Paths probability by dependency group}%
   \label{table:paths}
}
\end{floatrow}
\end{figure}

The $D^{2}T^{2}$ methodology, in its current formulation, offers an efficient strategy to deal with basic dependencies, intended as causal or stochastic relationships incurring between the basic events of a FT \cite{tolo2022integrated}.
\\The main procedure consists of:
\begin{enumerate}
    \item identifying dependent events within the FT and their associated dependency group;
    \item implementing dependency models capturing the overall dependency group relationship (unless dependency data is already available);
    \item analysing the dependency models to estimate joint metrics associated with the dependency groups (unless dependency data is already available);
    \item converting the initial FT into the corresponding BDD;
    \item integrating the joint metrics obtained into the BDD structure and extract the reliability information of interest.
\end{enumerate}
The first two steps are carried out on the basis of the available system knowledge as well as of analyst modelling assumptions. Any suitable technique can be adopted for the implementation of the dependency model, although current applications of the $D^{2}T^{2}$ approach have so far focused on the use of PNs and MMs models. These represent the dependency relationship, allowing to analyse its strength through the joint reliability metrics in output, as stated in step three. If deemed opportune, the results obtained can be compared with the assumption of independence in order to evaluate the significance of the dependency and provide evidence-based support for possible modelling assumptions or simplifications.
\\Taking into consideration the model in Fig. \ref{fig:bddft}, let assume the three basic events $\mathbf{X1}$, $\mathbf{X2}$ and $\mathbf{X5}$ to be dependent on each other and hence to belong to a dependency group $DG1={\mathbf{X1},\mathbf{X2},\mathbf{X5}}$. According to such assumption, the three corresponding components share the same load, with the portion assigned to $\mathbf{X2}$ and $\mathbf{X5}$ being fixed regardless of the state of the other components, while $\mathbf{X1}$ has the capability to process the entire load in the case of the unavailability of $\mathbf{X2}$ or/and $\mathbf{X5}$. However, under this condition, $\mathbf{X1}$ is forced to work beyond the design optimum, increasing the likelihood of malfunction and potentially triggering secondary failure. 
\\This setting can be represented by use of the PN model shown in Fig. \ref{fig:BEpn}. The numerical information required for its implementation is the input of the original FT model, with transitions F1, F2 and F6 in Fig.\ref{fig:BEpn} following the failure distributions assigned to basic events $\mathbf{X5}$, $\mathbf{X2}$ and $\mathbf{X1}$ respectively, while the repair distributions for the same become input to transitions R1, R2 and R3. Additional information is required for transitions F3, F4 and F5, modelling the the failure distribution of $\mathbf{X1}$ in case of additional load. If all components involved in the dependency group are characterised by constant rates (and thus have exponentially distributed failure and repair times), the PN model can be substituted with the MM in Fig.\ref{fig:BEmm}, allowing to reduce the computation demand of the analysis under the steady state assumption.
\\The fourth step is carried out using available conversion algorithms \cite{rauzy1993new}, obtaining the BDD structure from the initial FT model, as shown in the example in Fig.\ref{fig:bddft}.
\\In the presence of basic dependencies, as in the example under study, the probability of the identified BDD failure paths must be calculated according to the dependency groups entailed, as listed in Table \ref{table:paths}. 
In light of this, the corresponding top event probability for the FT in Fig.\ref{fig:bddft} can be calculated as:
\begin{equation}
    Q(G0) = \sum_{i=1}^{10} q(path_i)^0 q(path_i)^1
\end{equation}
For the computation of further reliability metrics, such as top event frequency or components importance measures, please refer to the algorithms presented in \cite{tolo2023fault}.

\subsection{Intermediate Events Dependency}
\label{sec:intermediateDep}
\begin{figure}
    \centering
    \includegraphics[width=\linewidth]{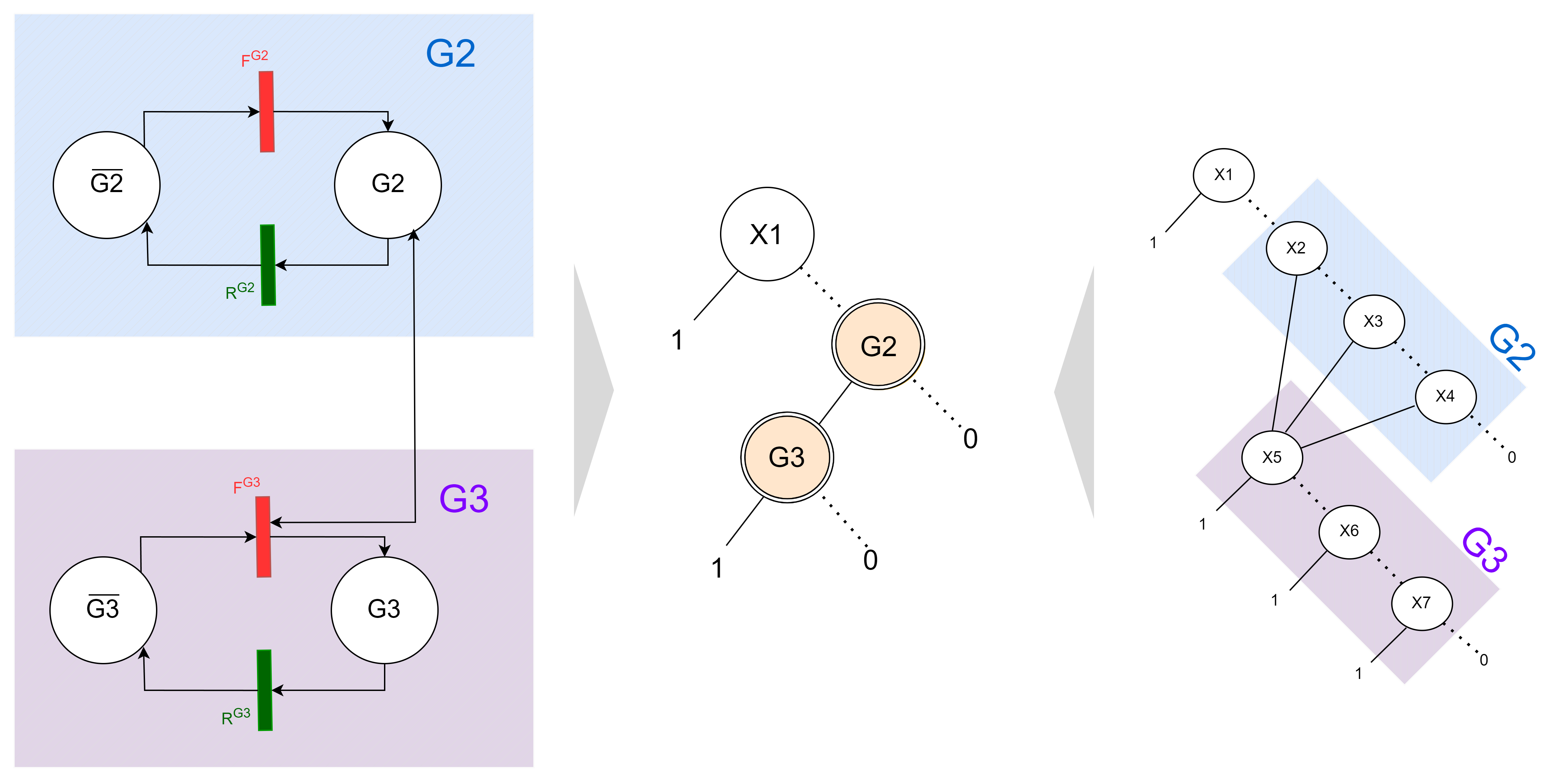}
    \caption{Dependency modelling for intermediate events G2 and G3}
    \label{fig:intermediate}
\end{figure}

In engineering systems, dependencies often occur between sets of components rather than individual ones: this can be triggered by redundancy (e.g., standby systems) \cite{clarke2010introduction}, common cause failure \cite{singh2021modeling}, asset management or maintenance strategies which involve entire batches of components \cite{sun2021group}. An example frequently found in common practice are emergency subsystem trains, whose components are frequently maintained, activated or brought to stand-by following coordinated procedures. From a modelling point of view, these dependencies still fall within the FT structure, but relate to intermediate events rather than basic ones.
\begin{figure}
    \centering
    \includegraphics[width=0.3\textwidth]{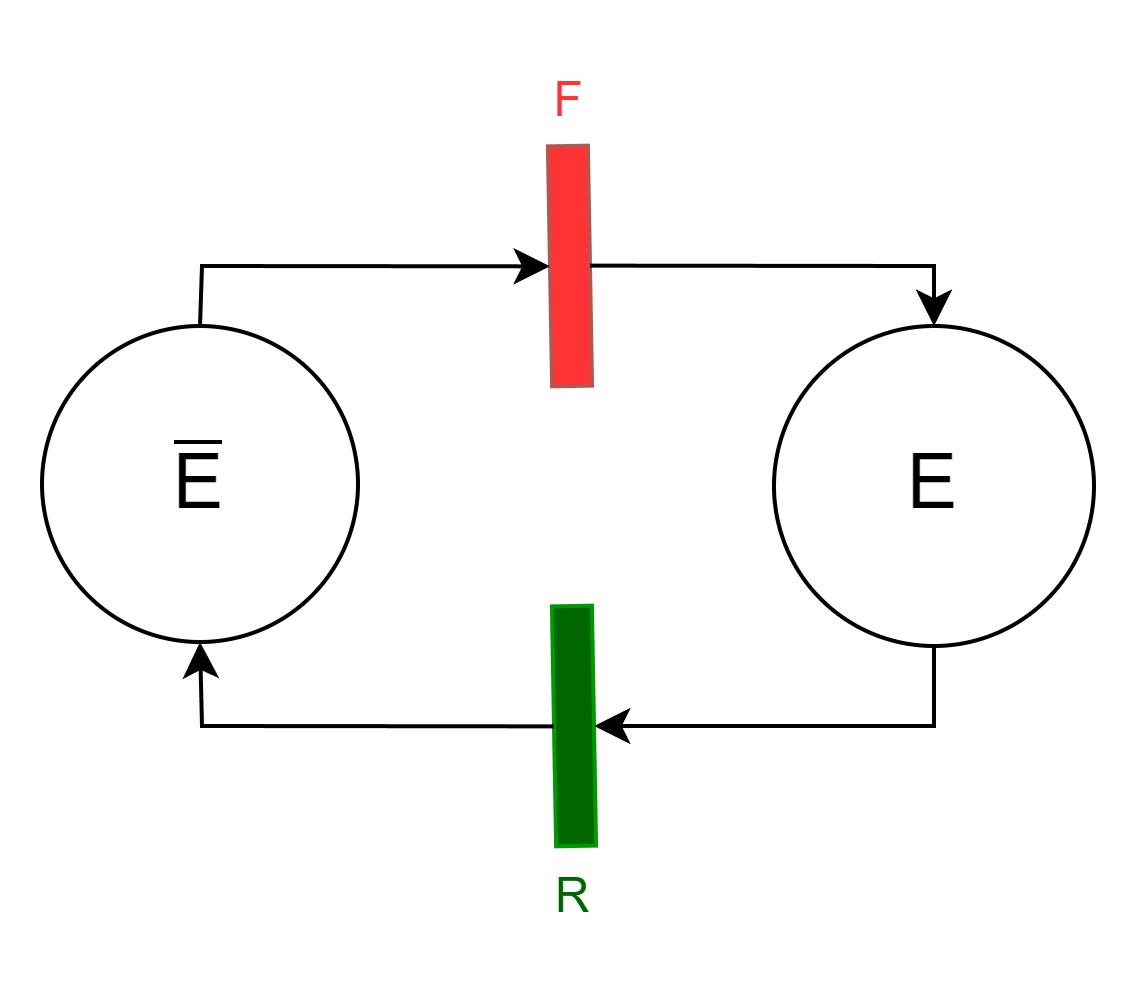}
    \caption{Overview of the generic dynamic model template for FT conversion}
    \label{fig:PNtemplate}
\end{figure}
\\The application of the original formulation of the $D^{2}T^{2}$ methodology to intermediate dependencies, implies the need to factor such relationships into the resulting basic dependencies existing between the underlying basic events (and hence components). This can put strain on the analyst and may result in convoluted support models, even for component trains of limited size.
\\In order to minimize the complexity of the dependency models, the $D^2T^2$ approach has been generalised to reduce intermediate dependencies to basic ones, subsequently resolved through the procedure discussed in Section \ref{sec:basicDep}. 
This is achieved by:
\begin{enumerate}
    \item Extracting the sub-trees associated with the dependent intermediate events within the FT;
    \item Extrapolating from these the reliability information required for the conversion of the sub-trees to dynamic sub-models;
    \item Combining the sub-models implemented in step 2 into the overall dynamic model characterizing the original dependency between intermediate events;
    \item Analysing the dynamic model extracting the joint reliability metrics for the dependent intermediate events;
    \item Substituting in the original FT structure equivalent complex events in place of the dependent intermediate events, characterized by the joint probability values obtained in step 4.
\end{enumerate}
For the purpose of converting individual sub-trees to dynamic sub-models (step 2), an universal PN template model is used, as shown in Fig.\ref{fig:PNtemplate} where the places labelled $\overline{E}$ and $E$ depict the non-occurrence and occurrence of a generic intermediate event $E$ respectively. The model allows to encode the reliability information encrypted in a generic sub-tree structure, including repair (transition $R$ in Fig.\ref{fig:PNtemplate}) as well as possible failure (transition $F$ in Fig.\ref{fig:PNtemplate}). In order to run the dynamic model, is then necessary to characterise the probabilistic distributions associated with these two transitions.
\\The probability distribution associated with transition $F$, models the shift of the intermediate event to a failure state. The related probability can be calculated from the analysis of the corresponding sub-tree. 
In absence of any repair procedure (e.g., if no reversible events are entailed by the sub-tree), only the failure transition $F$ is to be considered in the model and the associated probability equals the intermediate event reliability at any time value $t_i$, so that:
\begin{equation}
    F(t_i)=Q(E,t_i)
\end{equation}
where $F(t_i)$ refer to the value of the failure transition probability at time $t_i$, while $Q(E,t_i)$ indicates the unavailability of the generic intermediate event $E$ to be calculated with reference to the same time value.
Repeating this analysis of the sub-tree for n intervals of the mission domain, the empirical probability distribution associated with transition $F$ can be constructed interpolating the resulting support reliability/unavailability values:
\begin{equation}
    \mathbf{Q(E)}=\{ Q(E,t_0),Q(E,t_1),...,Q(E,t_n) \}
\end{equation}
\\If the intermediate event is to be considered reversible instead, the probability associated with transitions $F$ and $R$ in Fig.\ref{fig:PNtemplate} can be calculated on the basis of the time-dependent values of the failure and repair rates respectively, in turn obtained according to the following steps:
\begin{enumerate}
    \item Calculate the sub-tree top event probability and frequency over $n$ time intervals within the mission domain (by use of FT or $D^2T^2$ analysis, according to the presence of internal dependencies). This results in the identification of two vectors:
    \begin{equation}
    \label{eq:qandwvectors}
        \begin{split}
         \mathbf{Q(E)}&=\{Q(E,t_{0}),Q(E,t_{1}),...,Q(E,t_{n})\}\\
         \mathbf{W(E)}&=\{W(E,t_{0}),W(E,t_{1}),...,W(E,t_{n})\}
        \end{split}
    \end{equation}
where $Q(E,t_i)$ and $W(E,t_i)$ refer to the intermediate event E probability and frequency respectively computed at time $t_i$, while $t_{n}$ is equal to the overall mission time.
    \item Convert the probability and frequency values obtained in step 2 into failure rates over the time values considered, such as:
    \begin{equation}
    \label{eq:failRate}
        \lambda(t_{i})= \frac{W(E,t_{i})}{1-Q(E,t_{i})}\\
    \end{equation}
    where $i=0,...,n$, resulting in the failure rate vector:
    \begin{equation}
           \mathbf{\lambda}=\{\lambda(t_{0}),\lambda(t_{1}),...,\lambda(t_{n})\}
    \end{equation}
\item Convert the probability and frequency values obtained in step 2 into repair rates over the time values considered, such as:
    \begin{equation}
    \label{eq:repairRate}
       \nu(t_{i})= \frac{W(E,t_{i})}{Q(E,t_{i})}
    \end{equation}
    where i=0,...,n, resulting in the repair rate vector:
    \begin{equation}
    \label{eq:repairRate2}
           \mathbf{\nu}=\{\nu(t_{0}),\nu(t_{1}),...,\nu(t_{n})\}
    \end{equation}  
\end{enumerate}
The failure time distribution for transition $F$ can then be constructed on the basis of the rates values computed in step 3: if rates remain constant along the mission time (i.e., $\lambda(t_{i})=\lambda, \forall i=0,...,n$), the corresponding failure time can be considered exponentially distributed, with rate $\lambda$. Otherwise, a non parametric distribution is constructed over support probability values, namely $F(t_i)$. These indicate the probability of failure calculated at the time values considered, assuming the rates to remain constant over the interval (i.e., $\lambda(t_{i})=\lambda_i$). The empirical probability density function for the intermediate event occurrence can be then calculated from the interpolation of the resulting support points:
    \begin{equation}
       \label{eq:failSupportpdf}
         f(E,t_i)=\lambda(t_i)\cdot exp(-\lambda(t_i)\cdot (t_i))
    \end{equation}
    Finally, the cumulative distribution associated with transition $F$ is obtained interpolating the support points:
    \begin{equation}
         \label{eq:failSupport}
         F(t_i)=\sum_{j=0}^i f(E,t_j)
    \end{equation}
\\Similarly, the empirical probability distribution associated with the repair transition $R$ can be constructed from the interpolation of the cumulative probability distribution values:
    \begin{equation}
         \label{eq:repairSupport}
         R(t_i)=\sum_{j=0}^i r(E,t_j)
    \end{equation}
where $r(E,t_j)$ refer to the repair probability density function calculated from the repair rates estimated in step 4, so that:
\begin{equation}
       \label{eq:repairSupportpdf}
        r(E,t_i)=\nu(t_i)\cdot exp(-\nu(t_i)\cdot (t_i))
    \end{equation}
The degree of accuracy of the output probability distributions can be modified through the choice of the time intervals analysed: smaller intervals will better support the veracity of the rates constancy assumption, although increasing the computational effort of the analysis.
The PN template in Fig.\ref{fig:PNtemplate} can be extended further to include more failure mechanisms when more information is available, as discussed in the numerical application analysed in \cite{newNuclear}. However, case-specific modelling solutions are beyond the purpose of the current study.
\hfill
\\The dynamic models built for the intermediate event, can then be integrated in any dependency modelling entailing such event. 
\\For instance, let assume the FT in Fig.\ref{fig:bddft} to be characterised by a dependency involving the two intermediate events $\mathbf{G2}$ and $\mathbf{G3}$, triggered by the cold standby of $\mathbf{G3}$ which is activated only in the case of failure of $\mathbf{G2}$ (and hence can fail only if $\mathbf{G2}$ has failed.
\\First, the two PN sections representing the failure mechanisms of the two sub-trees must be obtained from the template in Fig.\ref{fig:PNtemplate}. With reference to the $\mathbf{G2}$ sub-tree in Fig.\ref{fig:bddft}, the probability associated with transition $F^{G2}$ is gathered from the analysis of the sub-tree $X2+X3+X4$, hence considering:
\begin{equation}
    \begin{split}
         \mathbf{Q(X2+X3+X4)}&=\{Q(X2+X3+X4,t_{0}),Q(X2+X3+X4,t_{1}),...,Q(X2+X3+X4,t_{n})\}\\
         \mathbf{W(X2+X3+X4)}&=\{W(X2+X3+X4,t_{0}),W(X2+X3+X4,t_{1}),...,W(X2+X3+X4,t_{n})\}
    \end{split}
\end{equation}
Calculating from these vectors the associated failure rates, according to Eq.\ref{eq:failRate}, the probability distribution for the transition $F^{G2}$ can be finally built on the basis of the resulting values as an exponential distribution, if the rates are constant in time, or as a parametric distribution with support points estimated according to Eq.\ref{eq:failSupport} if otherwise. Similarly, the probability distribution for transition $R^{G2}$ can be constructed on the basis of the repair rates obtained using Eq.\ref{eq:repairRate}.
\\The sub-net capturing the behaviour of intermediate event $\mathbf{G3}$ can be constructed analysing in a similar manner the sub-tree in order to define the associated transition $F^{G3}$ and $R^{G3}$.
\\Once the dependency model capturing the $\mathbf{G2}$ and $\mathbf{G3}$ dependency is completed, the substitution of the two sub-trees with the equivalent, homonymous basic events can be finalised, reducing the intermediate events dependency to a basic events one that can be resolved by the analysis of the reduced BDD (see Fig.\ref{fig:intermediate}) according to the original $D^2T^2$ procedure discussed in Section \ref{sec:basicDep}. 
\\This approach allows also for the consideration of dependencies internal to the sub-tree structures: if these embrace dependent events, the reduced sub-trees analysis resulting in the the reliability vectors in Eq.\ref{eq:qandwvectors}, as well as the combination of on demand events, needs to be carried out applying the $D^2T^2$ approach rather than using traditional FT techniques.

\begin{figure}
    \centering
    \includegraphics[width=0.7\linewidth]{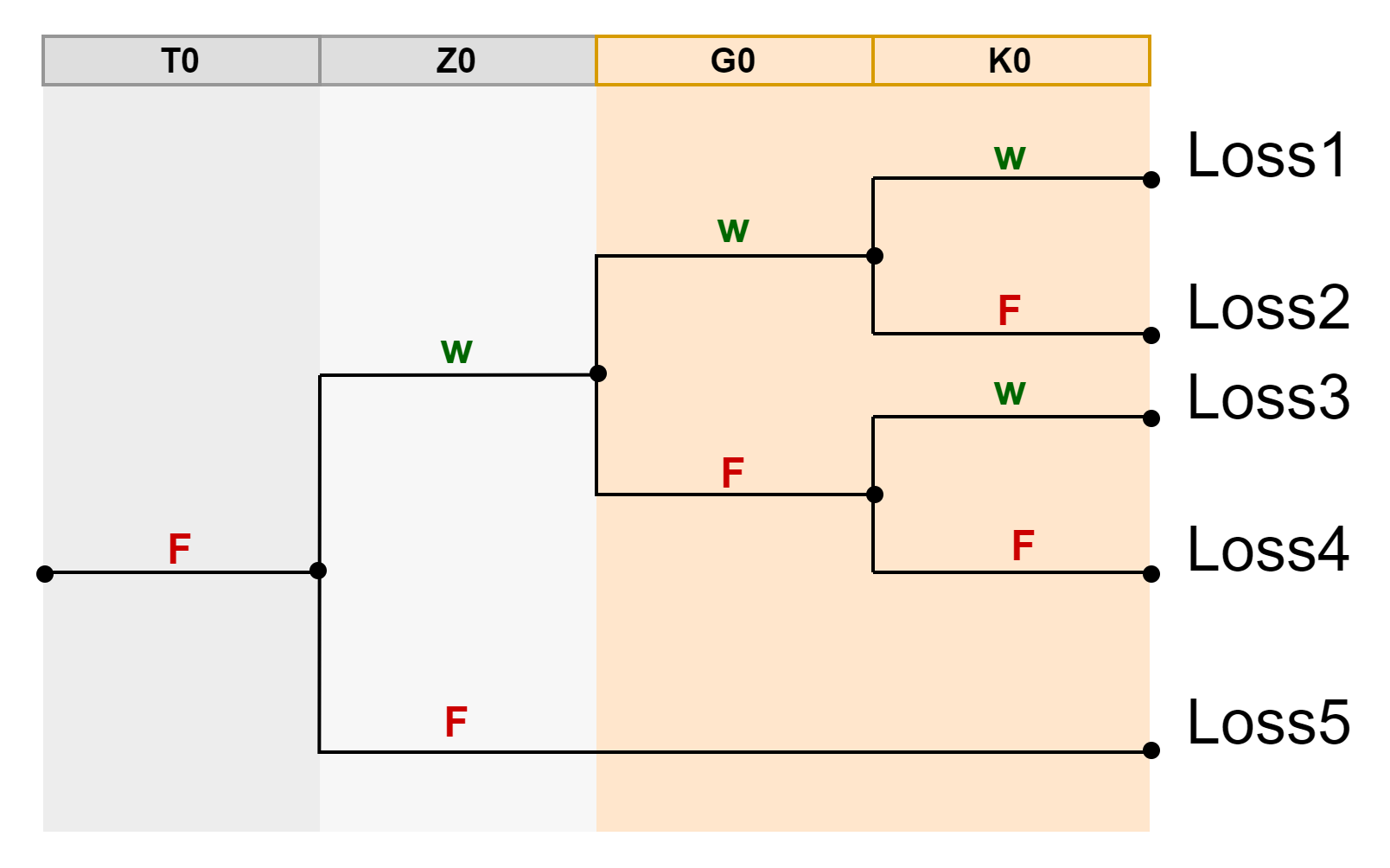}
    \caption{Example ET}
    \label{fig:et}
\end{figure}
\subsection{Fault Trees Dependency}
\label{sec:etDep}
\begin{figure}
    \centering
    \includegraphics[width=0.9\linewidth]{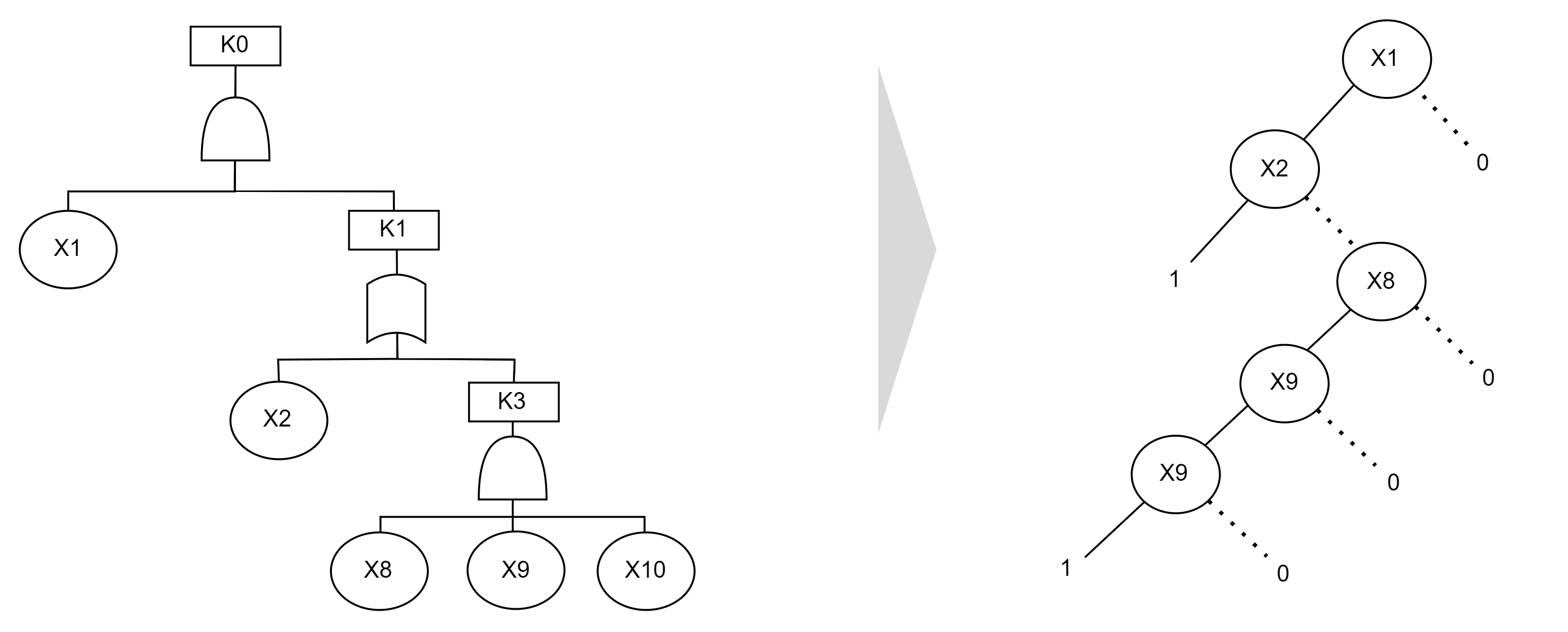}
    \caption{FT and correspondent BDD for top event K0}
    \label{fig:K0ft}
\end{figure}

\begin{figure}
\begin{subfigure}{.3\textwidth}
  \centering
  \includegraphics[width=\linewidth]{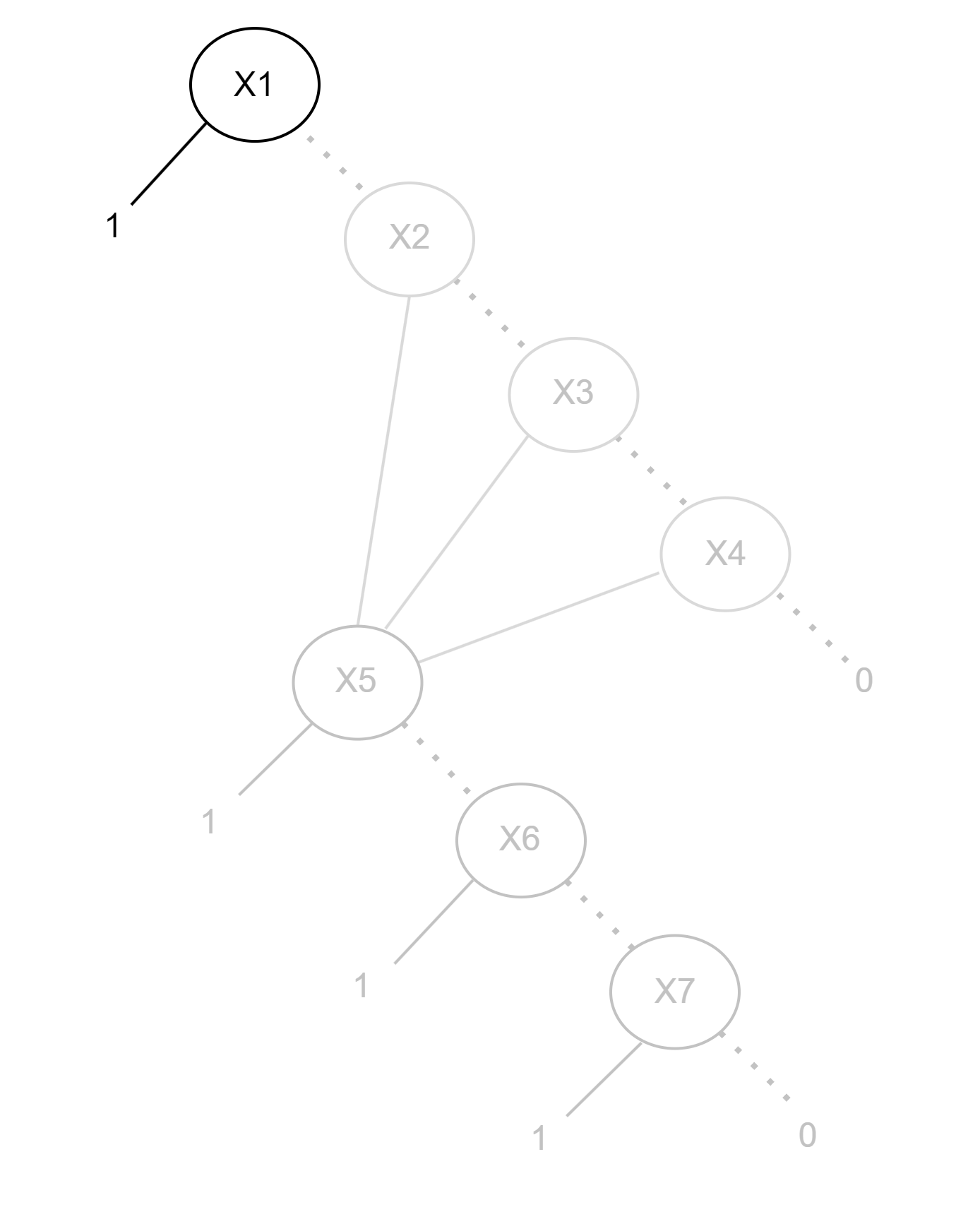}
  \caption{$q(G0,X1)$}
  \label{fig:G0X1}
\end{subfigure}%
\begin{subfigure}{.3\textwidth}
  \centering
  \includegraphics[width=\linewidth]{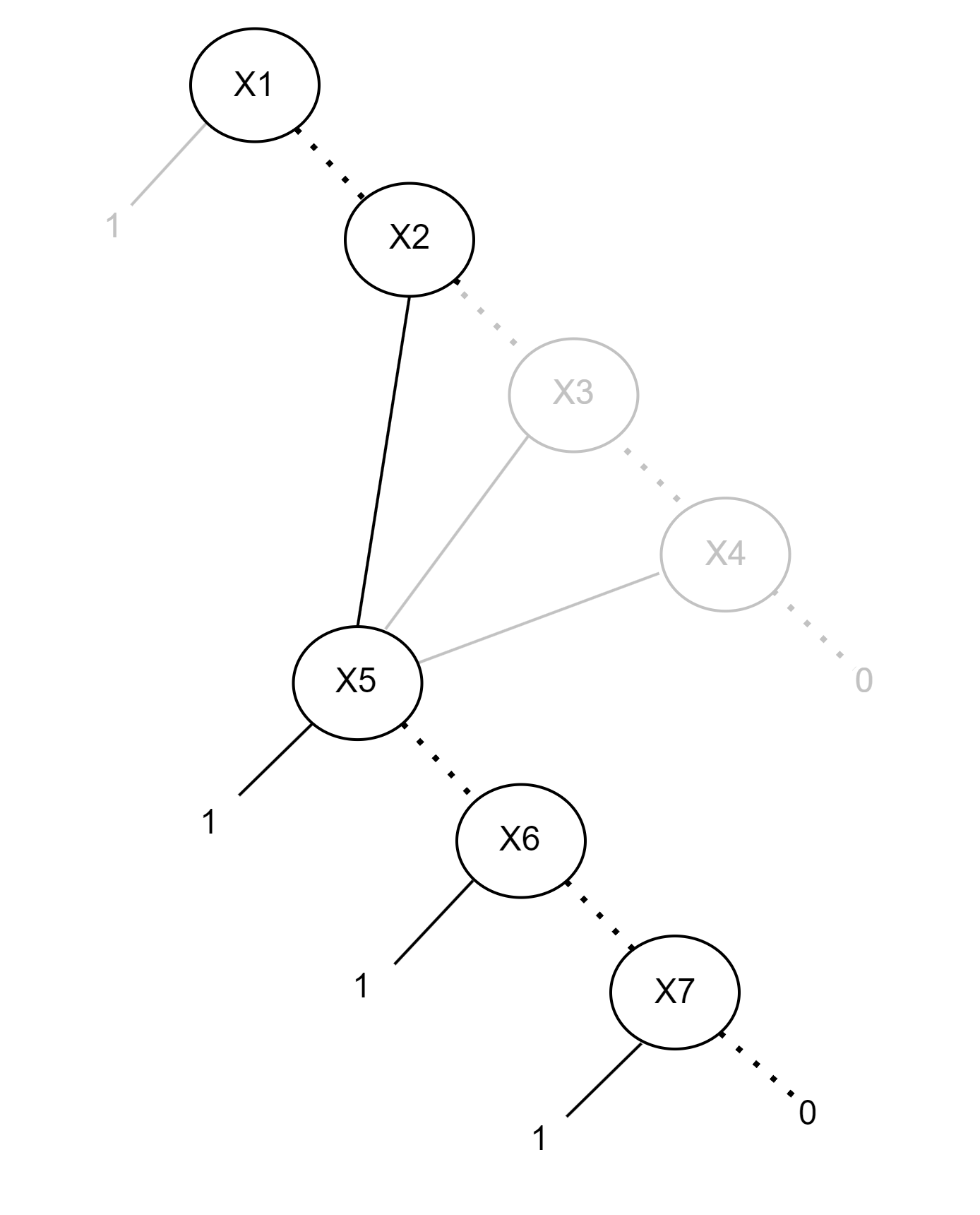}
  \caption{$q(G0,\overline{X1},X2$)}
  \label{fig:G0noX1X2}
\end{subfigure}
\begin{subfigure}{.3\textwidth}
  \centering
  \includegraphics[width=\linewidth]{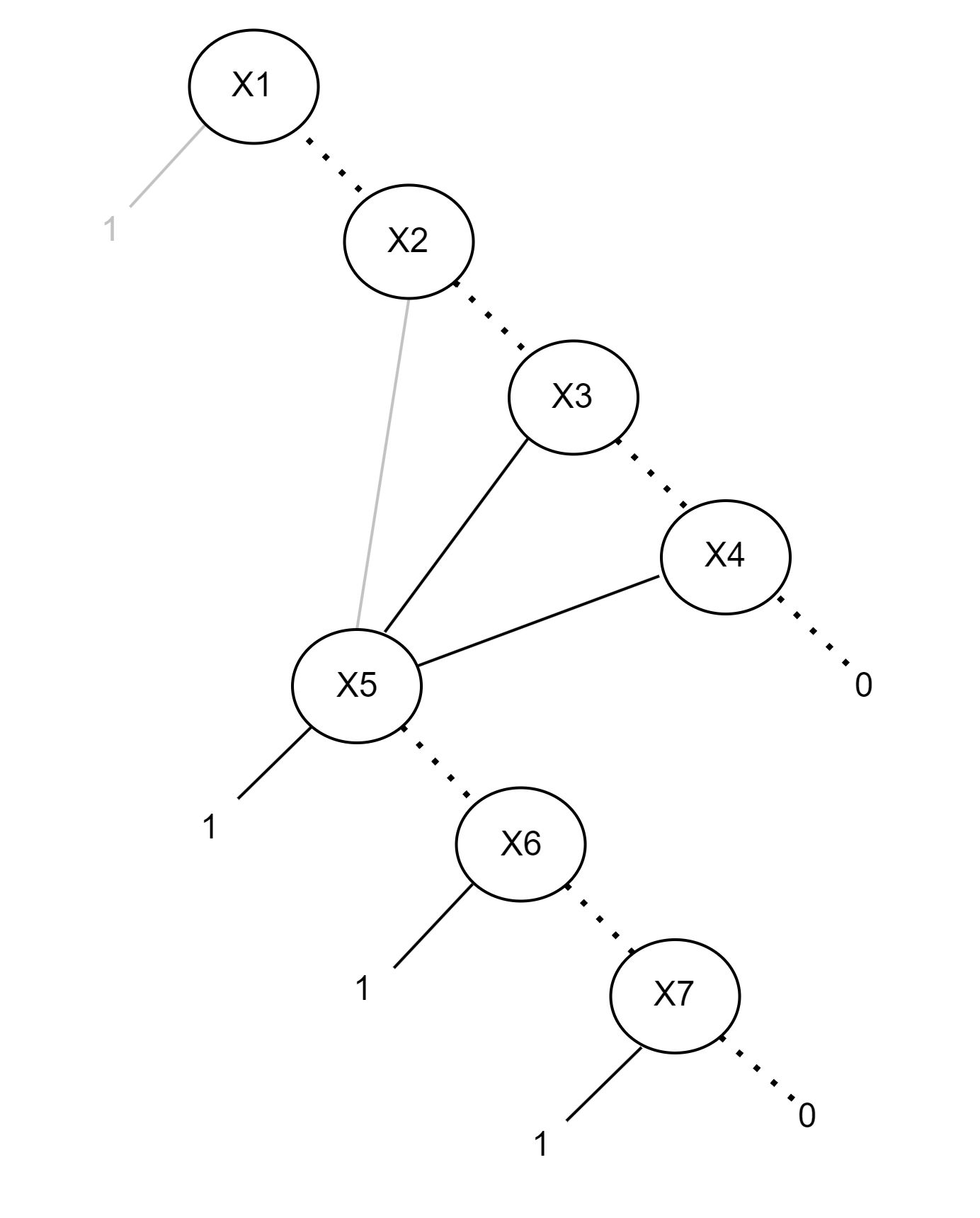}
  \caption{$q(G0,\overline{X1},\overline{X2}$)}
  \label{fig:G0noX1noX2}
\end{subfigure}
\caption{BDD encoding joint probabilities for FT G0}
\label{fig:G0bdds}
\end{figure}

Multiple subsystems (and hence their associated FTs) may be structurally dependent on each other, due for instance to shared resources or components. As discussed in Section \ref{sec:method}, such dependencies can be removed by instantiating their sources in the analysis of the individual BDD of the dependent FT models. In fact, as far as all sources of dependencies between these are explicitly instantiated, the resulting conditional top event probabilities are independent. 
This procedure can be applied to the computation of ETs including dependent FTs, allowing to take into account such dependencies in the analysis. 
\\With reference to the ET model in Fig.\ref{fig:et}, let assume FT $\mathbf{G0}$, shown in Fig.\ref{fig:bddft}, and $\mathbf{K0}$, shown in Fig.\ref{fig:K0ft}, to be dependent on each other due to the sharing of components $\mathbf{X1}$ and $\mathbf{X2}$. 
The frequency associated with the losses represented in the ET model must then be expressed as:
\begin{equation}
\label{eq:etfreq}
\begin{split}
W(Loss1)&=W(T0)q(\overline{Z0})q(\overline{G0},\overline{K0})\\
    W(Loss2)&=W(T0)q(\overline{Z0})q(\overline{G0},K0)\\
    W(Loss3)&=W(T0)q(\overline{Z0})q(G0,\overline{K0})\\
    W(Loss4)&=W(T0)q(\overline{Z0})q(G0,K0)\\
    W(Loss5)&=W(T0)q(Z0)
\end{split}
\end{equation}
The joint probabilities $q(\overline{G0},\overline{K0})$, $q(\overline{G0},K0)$ and $q(G0,\overline{K0})$ can be estimated from the conditioning of the $\mathbf{G0}$ and $\mathbf{K0}$ FT, instantiating the state of the dependency sources $\mathbf{X1}$ and $\mathbf{X2}$. As discussed in Section \ref{sec:BDD}, this is achieved through the BDD computation, identifying the failure paths relevant to the dependency to be instantiated.
\\With regards to the FT for event $G0$, this equals to split the analysis of the corresponding BDD into three sections: 
\begin{itemize}
    \item paths including $X1$ (i.e., the occurrence of $\mathbf{X1}$), highlighted in Fig.\ref{fig:G0X1}, from which it results:
    \begin{equation}
        q(G0|X1)=q(G0|X1,X2)=q(G0|X1,\overline{X2})=1
    \end{equation}
    \item paths including $\overline{X1}$ (i.e., non occurrence of $\mathbf{X1}$) and $X2$ (i.e., occurrence of $\mathbf{X2}$), highlighted in Fig.\ref{fig:G0noX1X2}. From this, the conditional probability can be expressed as:
    \begin{equation}        q(G0|\overline{X1},X2)=q(X5)+q(\overline{X5})q(X6)+q(\overline{X5})q(\overline{X6},X7)
    \end{equation}
    \item paths including the non occurrence of both $\mathbf{X1}$ and $\mathbf{X2}$ (i.e., $\overline{X1}$ and $\overline{X2}$), highlighted in Fig.\ref{fig:G0X1}, from which it results:
     \begin{equation}
     \begin{split}
q(G0|\overline{X1},\overline{X2})&=q(X3)q(X5)+q(X3)q(\overline{X5})q(X6)+q(X3)q(\overline{X5})q(\overline{X6},X7)\\
&+q(\overline{X3})q(X4)q(X5)+q(\overline{X3})q(X4)q(\overline{X5})q(X6)+q(\overline{X3})q(X4)q(\overline{X5})q(\overline{X6},X7)
    \end{split}
     \end{equation}
\end{itemize}

\begin{figure}
\begin{subfigure}{.4\textwidth}
  \centering
  \includegraphics[width=\linewidth]{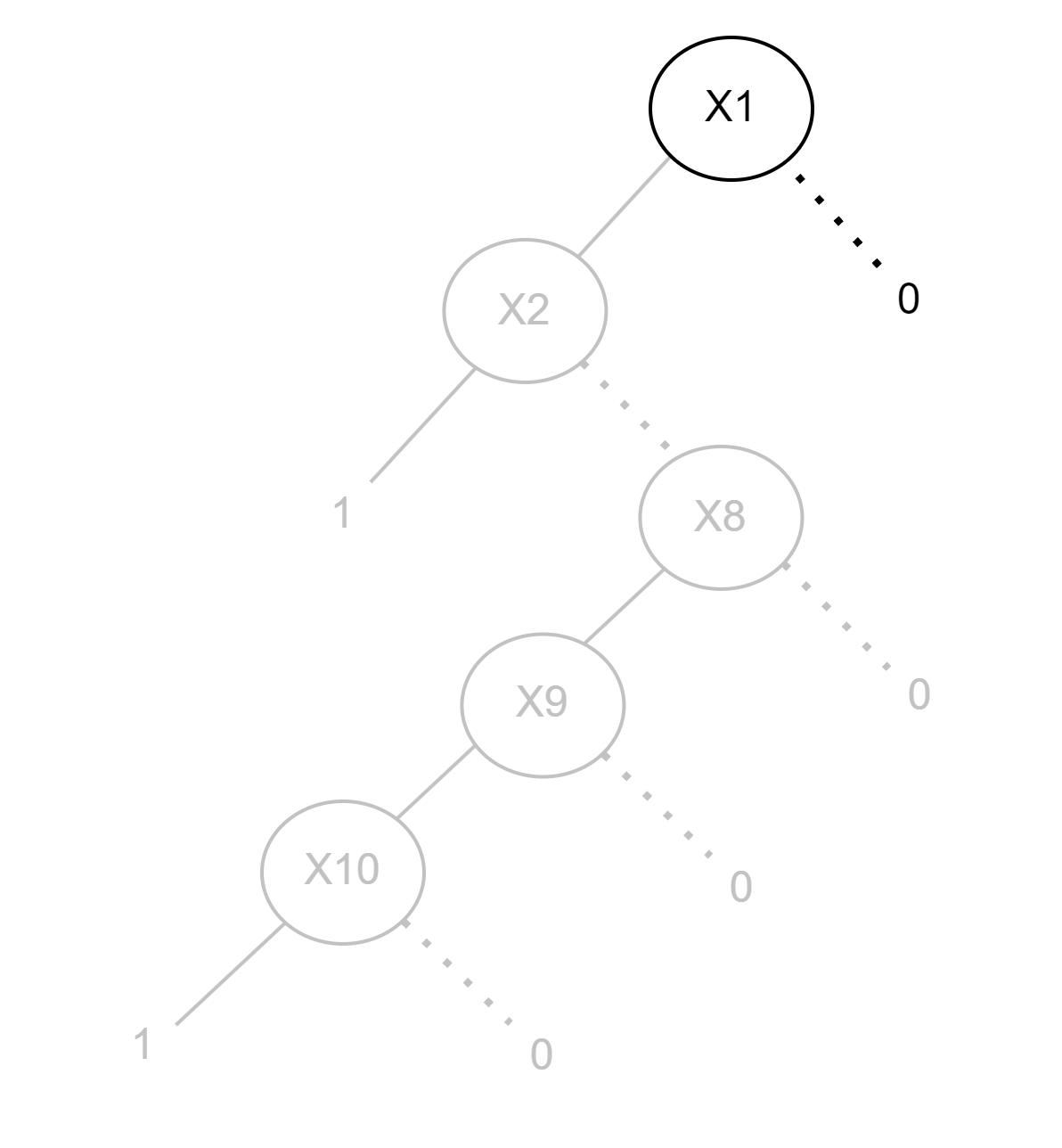}
  \caption{$q(K0,\overline{X1})$}
  \label{fig:K0noX1}
\end{subfigure}%
\begin{subfigure}{.4\textwidth}
  \centering
  \includegraphics[width=\linewidth]{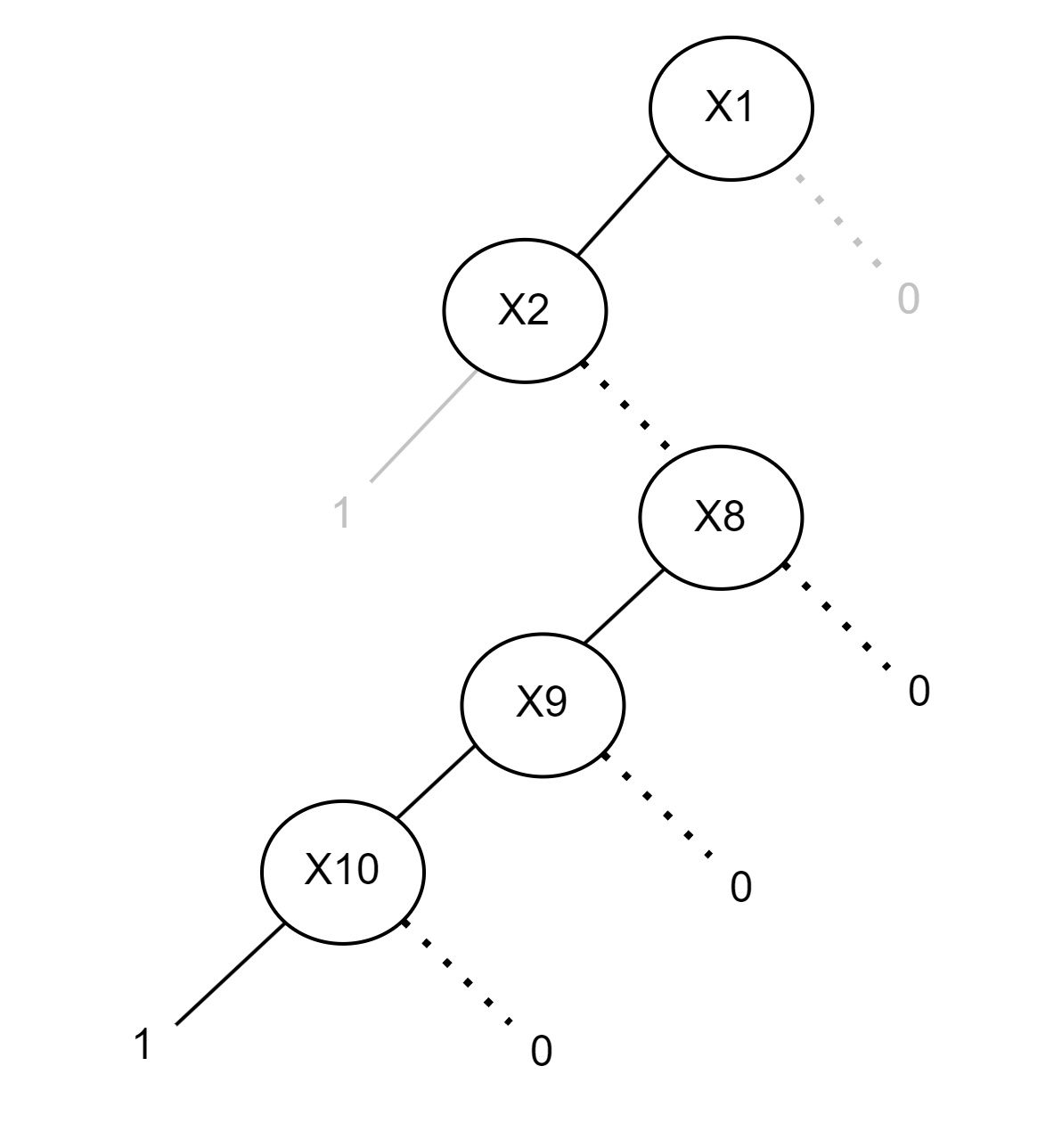}
  \caption{$q(K0,\overline{X1},\overline{X2})$}
  \label{fig:K0X1noX2}
\end{subfigure}
\begin{subfigure}{.4\textwidth}
  \centering
  \includegraphics[width=\linewidth]{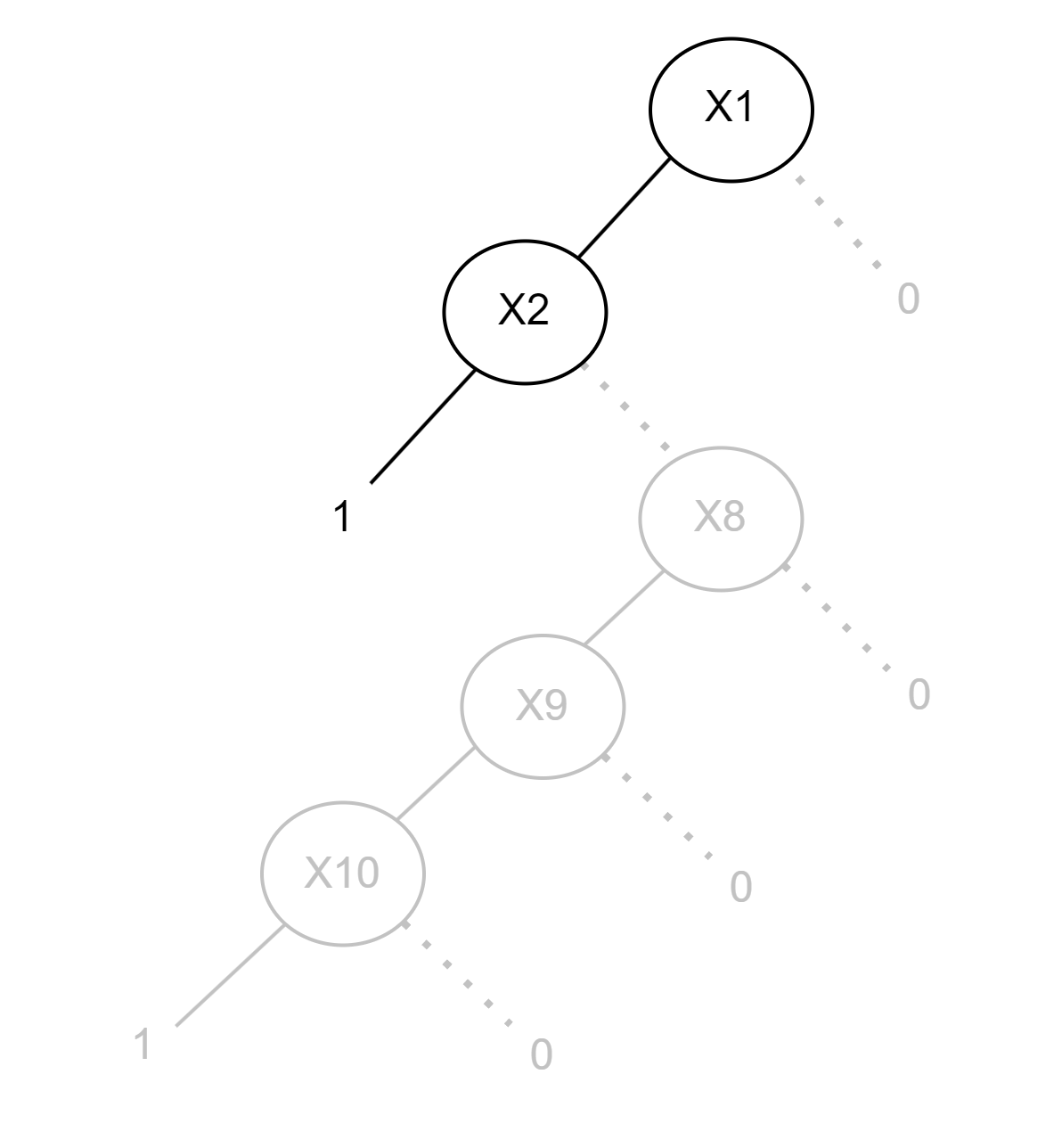}
  \caption{$q(K0,X1,X2)$}
  \label{fig:K0X1X2}
\end{subfigure}
\caption{BDD encoding joint probabilities for FT K0}
\label{fig:K0bdds}
\end{figure}
Similarly, the conditional probabilities of $\mathbf{K0}$ given the possible combination of states of $\mathbf{X1}$ and $\mathbf{X2}$ can be obtained:
\begin{itemize}
    \item analysing the BDD section highlighted in Fig.\ref{fig:K0noX1}, taking into account the non occurrence of $\mathbf{X1}$ and resulting in:
    \begin{equation}
q(K0|\overline{X1})=q(K0|\overline{X1},\overline{X2})=q(K0|\overline{X1},X2)=0
    \end{equation}
    \item analysing the BDD section highlighted in Fig.\ref{fig:K0X1noX2}, taking into account the occurrence of $\mathbf{X1}$ but not of $\mathbf{X2}$, so that:
    \begin{equation}
        q(K0|X1,\overline{X2})=q(X8)q(X9)q(X10)
    \end{equation}
    \item analysing the BDD section highlighted in Fig.\ref{fig:K0X1X2}, taking into account the occurrence of both events, $X1$ and $X2$, so that:
    \begin{equation}
        q(K0|X1,X2)=1
    \end{equation}   
\end{itemize}
The conditional probabilities obtained can be expressed in a vectorial form, such as:
\begin{equation}
     \begin{aligned}
    q(G0|\mathbf{X1},\mathbf{X2}) &= \begin{bmatrix}
           q(G0|X1,X2) \\
           q(G0|\overline{X1},X2) \\
           q(G0|X1,\overline{X2}) \\
           q(G0|\overline{X1},\overline{X2}) \\
         \end{bmatrix}
     \end{aligned}  
     \begin{aligned}
    q(K0|\mathbf{X1},\mathbf{X2}) &= \begin{bmatrix}
           q(K0|X1,X2) \\
           q(K0|\overline{X1},X2) \\
           q(K0|X1,\overline{X2}) \\
           q(K0|\overline{X1},\overline{X2}) \\
         \end{bmatrix}
     \end{aligned}  
\end{equation}
while the complementary probabilities can be computed from these as:
\begin{equation}
    \begin{split}
        q(\overline{G0}|\mathbf{X1},\mathbf{X2})&=1-q(G0|\mathbf{X1},\mathbf{X2})\\
        q(\overline{K0}|\mathbf{X1},\mathbf{X2})&=1-q(K0|\mathbf{X1},\mathbf{X2})\\
    \end{split}
\end{equation}
In light of this, the joint probabilities required to solve Eq.\ref{eq:etfreq} can be finally computed through the vectorial operations:
\begin{equation}
\label{eq:jointResults}
    \begin{split}
        q(\overline{G0},\overline{K0})&=\sum_i q(\overline{G0},\overline{K0}|\mathbf{X1},\mathbf{X2})_i=\sum_i [q(\overline{G0}|\mathbf{X1},\mathbf{X2})\odot q(\overline{K0}|\mathbf{X1},\mathbf{X2})]_i \\
        q(\overline{G0},K0)&=\sum_i q(\overline{G0},K0|\mathbf{X1},\mathbf{X2})_i=\sum_i [q(\overline{G0}|\mathbf{X1},\mathbf{X2})\odot q(K0|\mathbf{X1},\mathbf{X2})]_i\\
        q(G0,\overline{K0})&=\sum_i q(G0,\overline{K0}|\mathbf{X1},\mathbf{X2})_i=\sum_i [q(G0|\mathbf{X1},\mathbf{X2})\odot q(\overline{K0}|\mathbf{X1},\mathbf{X2})]_i\\
        q(G0,K0)&=\sum_i q(G0,K0|\mathbf{X1},\mathbf{X2})_i=\sum_i [q(G0|\mathbf{X1},\mathbf{X2})\odot q(K0|\mathbf{X1},\mathbf{X2})]_i
    \end{split}
\end{equation}
where $\odot$ indicates the element-wise vectorial product. Combining Eq.\ref{eq:jointResults} with Eq.\ref{eq:etfreq} allows to estimate the frequencies of the system losses modelled by the ET in Fig.\ref{fig:et}, completing the analysis.

\section{Conclusions}
A novel approach is proposed providing the systematic integration of dependency analysis in traditional Fault and Event Tree modelling framework. The methodology extends the current formulation of the Dynamic and Dependent Tree Theory, offering solutions for the modelling of intermediate event dependencies (e.g., between subsystem trains) as well as dependencies existing between Fault trees, through the integration of flexible modelling techniques such as Petri Nets and Markov Models. 
\\The defined methodology increases the overall flexibility of the Dynamic and Dependent Tree Theory original formulation, offering an exhaustive and systematic solution for the inclusion of dependencies affecting any level of traditional safety analysis modelling, regardless of the relationship type or location, while retaining the strengths of the original simulation framework.

\section*{Acknowledgement}
This work is funded by the Lloyd's Register Foundation, an independent global charity that helps to protect life and property at sea, on land, and in the air, by supporting high quality research, accelerating technology to application and through education and public outreach.

\end{document}